\documentclass[5p]{elsarticle}
 \usepackage{graphicx}
 \usepackage[font=footnotesize]{subfig} % subfig.sty for a double column floating figure using two subfigures

\usepackage{amssymb}
\usepackage{color}
\newcommand{\Omit}[1]{}

\def\beq{\begin{equation}}
\def\eeq{\end{equation}}
\def\bea{\begin{eqnarray}}
\def\eea{\end{eqnarray}}

\newcommand*{\eqref}[1]{Eq.~(\ref{eq:#1})}
\newcommand*{\eqlab}[1]{\label{eq:#1}}
\newcommand*{\figref}[1]{Fig.~\ref{fig:#1}}
\newcommand*{\figlab}[1]{\label{fig:#1}}

\newcommand*{\secref}[1]{Section~\ref{sec:#1}}
\newcommand*{\seclab}[1]{\label{sec:#1}}

\def\VYP#1#2#3{{\bf #1}, #3 (#2)}  % Volume, page (Year)

\def\PL#1#2#3{Phys.~Lett.~\VYP{#1}{#2}{#3}}

\begin{document}

\begin{frontmatter}



\title{Macroscopic Geo-Magnetic Radiation Model;
Polarization effects and finite volume calculations.}


\author[KVI]{Krijn D. de Vries}
\ead{dvries@kvi.nl}
%\author[KVI]{Ad M. van den Berg}
\author[KVI]{Olaf Scholten}
\author[SUBA]{and Klaus Werner}
\address[KVI]{Kernfysisch Versneller Instituut, University
of Groningen, 9747 AA, Groningen, The Netherlands}
\address[SUBA]{SUBATECH,
University of Nantes -- IN2P3/CNRS-- EMN,  Nantes, France}

\begin{abstract}
An ultra-high-energy cosmic ray (UHECR) colliding with the Earth's atmosphere gives rise to an Extensive Air Shower (EAS). Due to different charge separation mechanisms within the thin shower front coherent electromagnetic radiation will be emitted within the radio frequency range. A small deviation of the index of refraction from unity will give rise to Cherenkov radiation up to distances of 100 meters from the shower core and therefore has to be included in a complete description of the radio emission from an EAS. Interference between the different radiation mechanisms, in combination with different polarization behavior will reflect in a lateral distribution function (LDF) depending on the orientation of the observer and a non-trivial fall-off of the radio signal as function of distance to the shower core.

\end{abstract}

\end{frontmatter}

% main text

\section{Introduction}
Recent results to detect radio emission from UHECR induced particle cascades at the LOPES~\cite{Fal05,Ape06} and CODALEMA~\cite{Ard06,Ard09} experiments triggered plans to install a large array of radio antenna's at the Pierre Auger Observatory~\cite{Ber07,Cop09,Rev09}. Both experiments clearly indicate that the main emission mechanism is induced by the deflection of charged particles in the Earth magnetic field. The emission process can be described in a microscopic approach as done in for example~\cite{Hue08}, where the field is calculated from the track of a single particle and a summation is done over all particles. A different approach is given by a macroscopic description where the coherent emission from induced macroscopic currents and charges in the shower front is calculated. This idea is investigated since the earliest days of radio detection from EAS~\cite{Jel65,Por65,Kah66,All71}. Recently this approach is implemented using realistic shower profiles by the Macroscopic Geo-Magnetic Radiation (MGMR)~\cite{Sch08,Wer08,KdV10} model. Recent results from comparison between the MGMR and REAS3~\cite{Arena31} simulations for the first time showed good agreement between two completely different models~\cite{Arena34}.

Due to the interest in detecting radio emission from cosmic ray induced particle cascades on a large scale at the Pierre Auger Observatory~\cite{Ber07}, it is of importance to understand the underlying emission mechanisms and their lateral behavior. In~\cite{KdV10} the two most important emission mechanisms, geomagnetic radiation and radiation due to a net negative charge-excess in the shower front also known as the Askaryan effect, are discussed. It follows that the LDF depends strongly on the geometry considered and one has to take into account the observer angle dependency. In section~\secref{model} we will give a short review of the MGMR model and the different radio emission mechanisms. Interference effects and polarization patterns for the different mechanisms will be discussed in~\secref{Pol}. The consequences of these effects for the LDF will be discussed in~\secref{LDF}. Finally the effect due to the small deviation of the index of refraction from unity leading to Cherenkov radiation will be discussed in~\secref{cher}.

\section{The model\seclab{model}}
The MGMR model describes the emission due to the variation of a net macroscopic current in the shower front. This current, as follows from Monte-Carlo simulations, can be described using three macroscopic distributions. The total number of particles in the thin shower front at a fixed negative emission time ($t'<0$), $f_t(t')$. The longitudinal distribution of particles in the shower front, $f_p(h)$, where $h$ is the distance in meters behind the imaginary shower front traveling with the speed of light toward the surface of the Earth, and the lateral distribution of particles within the shower front which is neglected at the moment. It follows that the pulse shape and height contain direct information about these macroscopic distributions~\cite{Ar-Sch09}.

The shower front will be located at a height $z=-\beta t'$, defined in such a way that $t'=0$ when the shower hits the surface of the Earth. Hence the total number of particles corresponding to the negative emission time $t'$ at a height $z=-\beta t'+h$ is given by,
\beq
N(z,t')=N_e f_t(t')\,f_p(h)\;,
\eeq
where $N_e$ is the total number of particles in the shower at the shower maximum. The induced current in the pancake is now given by,
\beq
j^{\mu}(x,y,z,t')= v^{\mu} \, q \, e\, N(z,t') \delta(x) \delta(y)\;,
\eqlab{current}
\eeq
where $q$ is the charge of the particles considered, $v^{\mu}$ the four-velocity, and the $\delta(x) \delta(y)$ term takes care of the lateral distribution of the particles in the shower front, hence all particles are projected along the shower axis.   

The electric field is now obtained through the Li\'enard Wiechert potentials from classical electrodynamics~\cite{Jac-CE},
\beq
A^{\mu}(\vec{x},t)=\frac{j^{\mu}(\vec{x},t')}{{\cal D}}|_{t'=t_r},
\eqlab{Vect-pot}
\eeq
to be evaluated at the retarded time $t'=t_r$ given by,
\beq
ct_r=\frac{ct-n^2\beta_s h-n{\cal D}} {(1-n^2\beta_s^2)} \;\eqlab{ret}.
\eeq
Here $n$ is the index of refraction of air, $t$ is the observer time, and ${\cal D}$ is the retarded distance given by,
\beq
\eqlab{ret-dist}
{\cal D} = \sqrt{(-c\beta_s t+h)^2+(1-n^2\beta_s^2)d^2}  \;.
\eeq
The electric field is now obtained through the standard relation,
\beq
\vec{E}(\vec{x},t)=-\frac{d}{d\vec{x}}A^{0}(\vec{x},t)-\frac{d}{d(ct)}\vec{A}(\vec{x},t)
\eqlab{E-field}
\eeq
\section{Polarization and Interference}\seclab{Pol}
In~\cite{Sch08}, and~\cite{Wer08} several different emission mechanisms are discussed. We will focus on the three strongest contributions. The geomagnetic emission mechanism due to the deflection of electrons and positrons in the Earth magnetic field, gives rise to a net current in the direction of the Lorentz force acting on these particles. A secondary moving dipole contribution due to the geomagnetic charge separation inside the shower front, and the emission due to a net charge-excess in the shower, also known as the Askaryan effect~\cite{Ask62}. 

We study the ground polarization pattern for a perpendicular incoming air shower with the magnetic field pointing to the North. Since for the geomagnetic contribution an equal amount of electrons and positrons is assumed, only the vector potential in the direction of the induced drift velocity survives since the electrons and positrons are deflected in opposite directions in the Earth magnetic field. From~\eqref{E-field} it follows immediately that the signal is, independent of observer position, fully polarized in the $\hat{n}=-\vec{e}_{\beta}\times\vec{e}_{B}$ direction, where $e_{\beta}$ is the unit vector pointing along the shower axis, and $e_{B}$ the unit magnetic field vector~\figref{ground_pat} (top).

The zero'th component due to the net charge-excess in the shower front gives rise to a radial polarization pattern shown in~\figref{ground_pat} (bottom)~\cite{Wer08,KdV09}. The $xy$-polarization pattern, projected on the ground plane, for the moving dipole contribution is given by a typical dipole pattern. 
\begin{figure}[!ht]
\centerline{
\includegraphics[width=.48\textwidth, keepaspectratio]{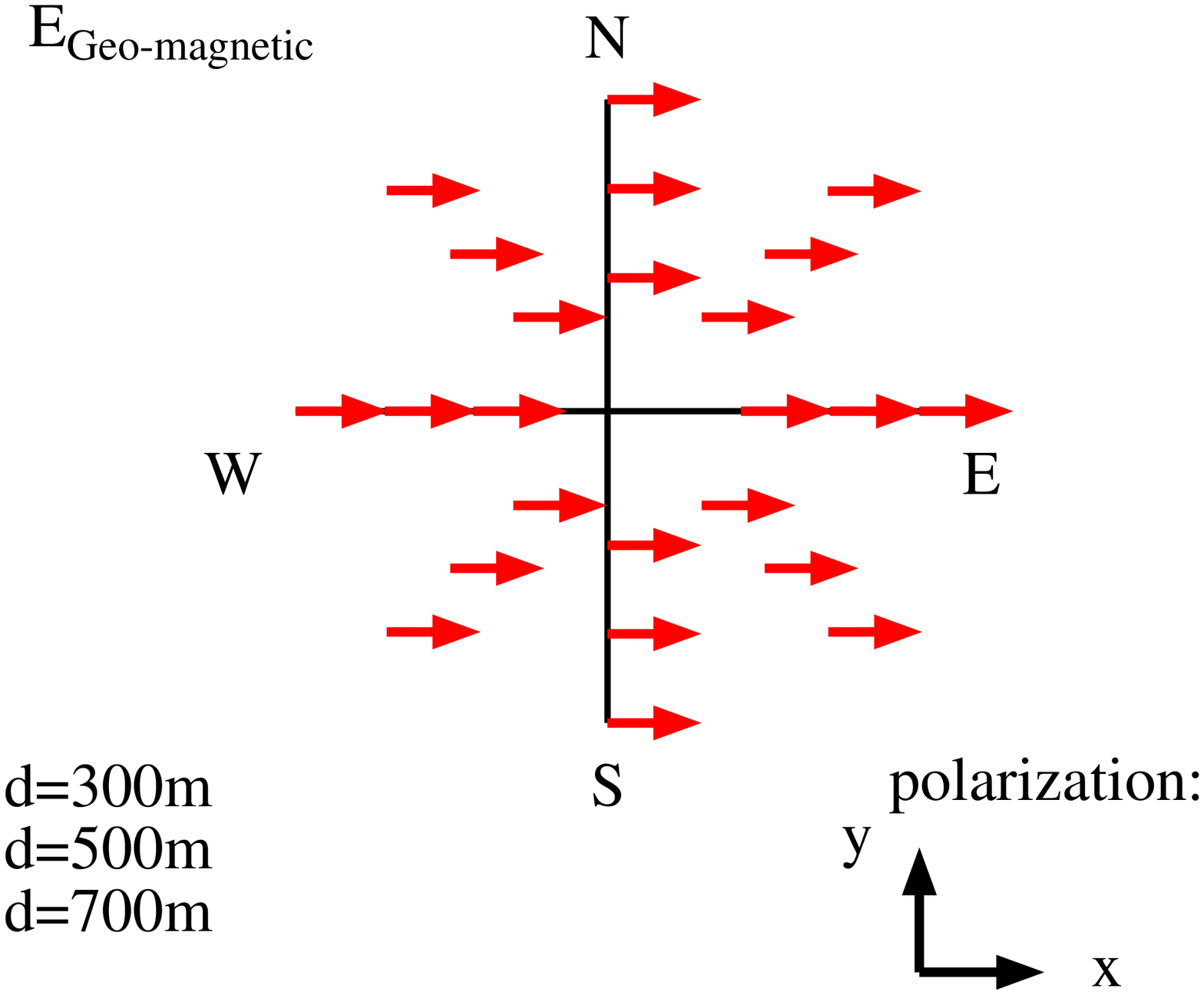}}
\centerline{
\includegraphics[width=.48\textwidth, keepaspectratio]{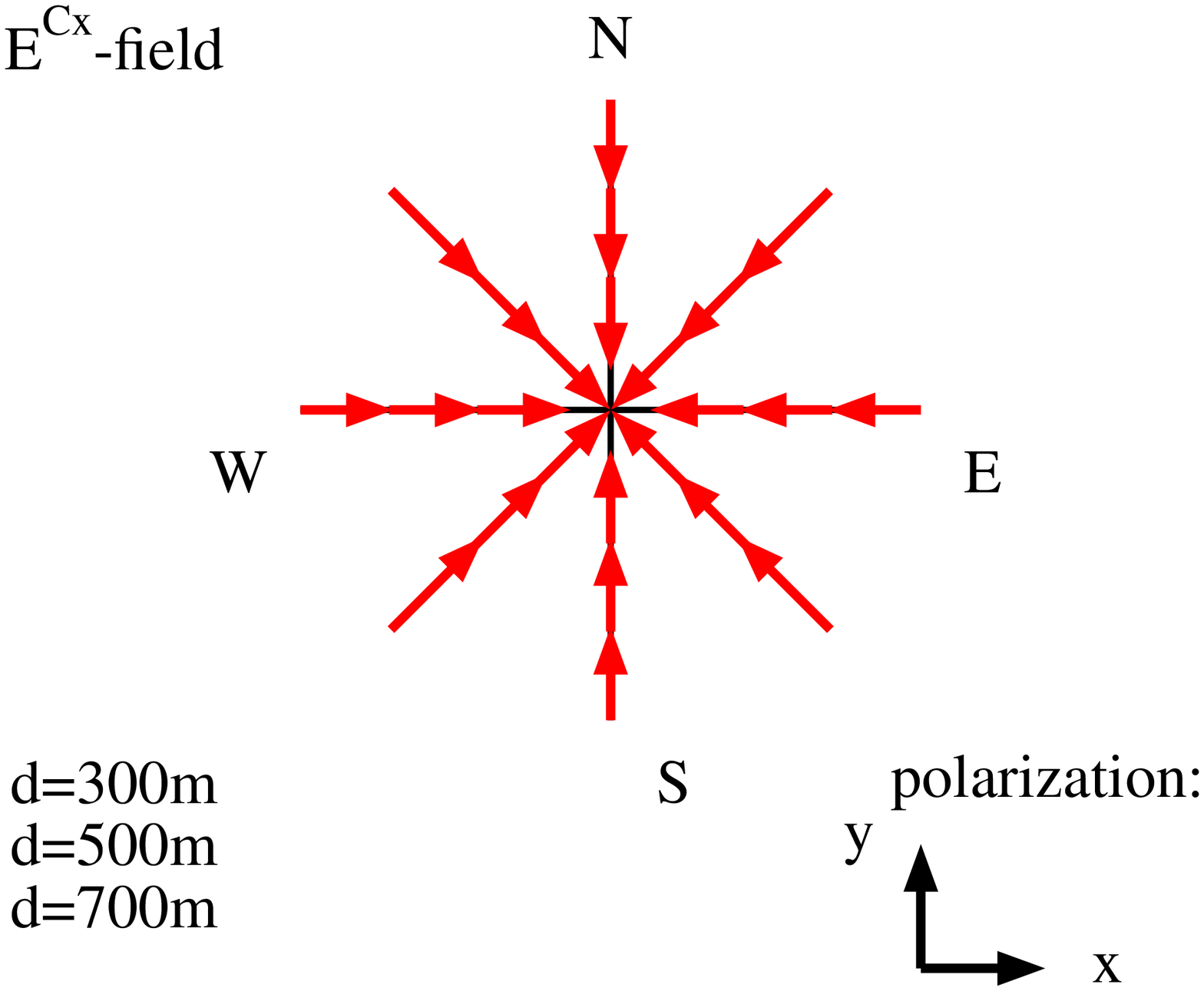}}
\caption{The polarization in the $xy$-plane seen by an observer at different observer positions w.r.t. the impact point of the shower for the geomagnetic contribution (top), and the charge-excess contribution (bottom).}
\figlab{ground_pat}
\end{figure}
A direct consequence of the different polarization patterns is different interference behavior as function of observer position. In the case that the geomagnetic and moving dipole polarization are pointing to the East, an observer placed West from the shower core will see constructive interference between the charge-excess and geomagnetic component in the East-West polarization and see nothing in the North-South polarization. When the observer is placed to the North of the shower core, the geomagnetic field will still be seen in the East-West polarization, whereas the charge-excess field is pointing in the North-South direction and no interference will take place. An observer placed to the East of the shower core will again see no field in the North-South polarization, nevertheless the charge-excess and geomagnetic components on this side of the shower core will interfere destructively. These interference effects as function of observer position can give a great tool to disentangle and study the different emission mechanisms~\cite{Sch10}. 

\section{Radial dependence of the radio signal}\seclab{LDF}
As discussed in the previous section, the interference between different contributions is strongly observer position dependent. As a direct consequence one has to be extremely careful in determining an LDF~\cite{Lop10}. For different observer positions constructive or destructive interference can take place and the field strength will, depending on the geometry, have strong fluctuations. 

In~\figref{ldf} the LDF for the three different contributions is shown for the maximum of the unfiltered electric field strength. It can be seen that the LDF's for the geomagnetic contribution and the dipole contribution follow each other, whereas the LDF for the charge-excess contribution is less steep and at distances larger than approximately $500$~m, the charge-excess contribution becomes the leading contribution. The LDF at a fixed observer angle leading to destructive interference between the three contributions is also shown in this plot. It follows that even for a fixed observer angle one has to be extremely careful since the LDF falls off less steep for larger distances than the distance where maximum destructive interference occurs.
\begin{figure}[!ht]
\centerline{
\includegraphics[width=.48\textwidth, keepaspectratio]{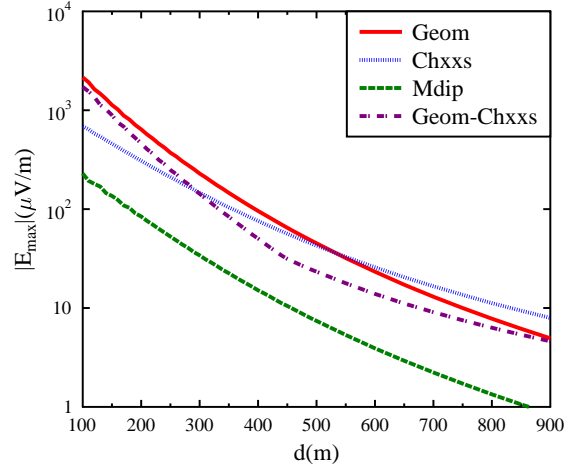}}
\caption{The LDF for the maximum of the unfiltered electric field strength for the three different emission mechanisms, the geomagnetic emission, dipole emission and the emission due to a net charge-excess in the shower also known as the Askaryan effect. The LDF for destructive interference of the charge-excess radiation and geomagnetic radiation illustrates that one has to be extremely careful by determining a LDF.}
\figlab{ldf}
\end{figure}

\section{Cherenkov radiation in EAS.}\seclab{cher}
Since the index of refraction in air slightly deviates from unity $O(10^{-4})$, even-though this deviation is small, Cherenkov effects can occur due to extremely high-energetic particles in the shower front. Mathematically, the retarded distance defined in~\eqref{ret-dist} can become zero leading to a singularity in the vector potential. In general, this will occur for the contribution to the current at a critical distance $h=h_c$ behind the shower front. From~\eqref{ret-dist} a simple expression for this critical distance can be derived.
\beq
h_c=\beta ct+\sqrt{n^{2}\beta^2-1}\;d,
\eeq
It follows that if the distance $d$ of the observer from the shower axis increases, $h_c$ increases in such a way that the angle between the shower axis and the vector pointing from the emission point to the observer, the Cherenkov angle, stays constant. Another way to visualize this is to look at the retarded shower time as a function of the observer time. This plot is shown in~\figref{trett} for the cases of $n=1$ and $n=1.0003$, and $n=n(z)$ where the index of refraction is modeled to depend on the shower height. For $n=1$, the signal which is emitted at large negative retarded shower times $t_r$, corresponding to large heights $z=-\beta t_r+h$, will travel along with the same speed as the shower and arrive at the antenna approximately at the same time of the shower hitting the ground, whereas a signal emitted at later stages has to cover the larger lateral distance toward the antenna and arrives at later times. For a fixed observer time $t$, an infinitesimal part of the shower will contribute to the observed signal leading to a finite contribution to the electric field. For the $n=1.0003$ case however, the signal emitted at large negative retarded shower times travels slightly slower than the shower itself, hence there will be an optimal height from where the signal reaches the antenna first, and as can be seen from~\figref{trett}, since the derivative of the negative shower time diverges at this point, a finite part of the shower will contribute to this signal giving it a 'boost' which is seen as Cherenkov radiation. 
\begin{figure}[!ht]
\centerline{
\includegraphics[width=.48\textwidth, keepaspectratio]{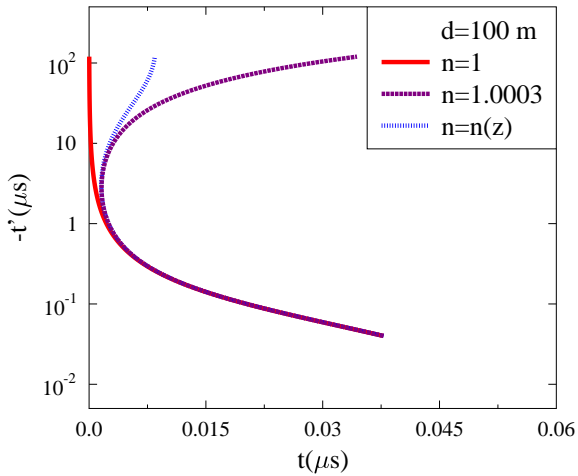}}
\caption{The negative emission time $-t'$ as function of the observer time $t$ for three different values of the index of refraction.}
\figlab{trett}
\end{figure}
This 'boost' will not only be seen from what is normally considered as Cherenkov radiation, the radiation from a net charge moving faster than the speed of light in the medium. Even if there would be no net charge inside the shower front, which in our case would correspond to geomagnetic radiation, the sideward induced current still moves faster than the speed of light in the medium, and hence the electric field due to geomagnetic radiation gets 'boosted'.

To see the importance of this effect, one needs to know the typical ground distance where one still observes Cherenkov radiation. From Monte-Carlo simulations it follows that most of the particles are located close to the shower front, hence $h=0$. From~\eqref{ret-dist} and~\eqref{ret}, an expression can be found for the observer distance where the Cherenkov emission is observed as function of emission height,
\bea
d_c&=&\sqrt{n^2\beta^2-1}\;z_c.
\eea
In the realistic case where the shower reaches its maximum at $4$~km height this critical distance becomes $d_c\approx 100$~m. This means that already at intermediate distances Cherenkov emission cannot be ignored and has to be included into realistic electric field calculations.

\section{Summary and Conclusions}
There are several emission mechanisms that give rise to radio emission from EAS. We discussed the three largest contributions, the geomagnetic contribution, the contribution due to a net charge-excess in the shower front, and the contribution due to moving dipole. It follows that all three contributions show different polarization patterns as function of observer position, giving an excellent tool to distinguish between the different emission mechanisms. The interference effects also have important consequences for the LDF, since constructive and destructive interference at different observer positions lead different field strengths. A first introduction to Cherenkov radiation from EAS is given, where it is shown that although the index of refraction deviates only slightly from unity, Cherenkov emission will be seen at intermediate distances from the shower core and cannot be neglected. It is discussed that this emission will not only come from a net charge moving faster than the speed of light in the medium. In the case of geomagnetic radiation there will be an induced current which itself travels faster than the speed of light in the medium also giving rise to Cherenkov effects.

\section{Acknowledgment}
This work is part of the research program of the `Stichting voor Fundamenteel
Onderzoek der Materie (FOM)', which is financially supported by the `Nederlandse
Organisatie voor Wetenschappelijk Onderzoek (NWO)'.

\end{document}